\documentclass[12pt]{iopart}
\begin{document}

\title[]{Non-isospectrality of the generalized Swanson Hamiltonian and harmonic oscillator}

\author{Bikashkali Midya, P P Dube and Rajkumar Roychoudhury}

\address{Physics and Applied Mathematics Unit, Indian Statistical Institute, Kolkata-700108, India.}
\ead{bikash.midya@gmail.com, ppdube1@gmail.com, raj@isical.ac.in}
\begin{abstract}
The generalized Swanson Hamiltonian $H_{GS} = w (\tilde{a}\tilde{a}^\dag+ 1/2) + \alpha \tilde{a}^2 + \beta \tilde{a}^{\dag^2}$ with $\tilde{a} = A(x)d/dx + B(x)$, can be transformed into an equivalent Hermitian Hamiltonian with the help of a similarity transformation. It is shown that the equivalent Hermitian Hamiltonian can be further transformed into the harmonic oscillator Hamiltonian so long as $[\tilde{a},\tilde{a}^\dag]=$ constant. However, the main objective of this paper is to show that though the commutator of $\tilde{a}$ and $\tilde{a}^\dag$ is constant, the generalized Swanson Hamiltonian is not necessarily isospectral to the harmonic oscillator. Reason for this anomaly is discussed in the frame work of position dependent mass models by choosing $A(x)$ as the inverse square root of the mass function.

\end{abstract}

%Uncomment for PACS numbers title message
%\pacs{00.00, 20.00, 42.10}
% Keywords required only for MST, PB, PMB, PM, JOA, JOB?
%\vspace{2pc}
%\noindent{\it Keywords}: Article preparation, IOP journals
% Uncomment for Submitted to journal title message
%\submitto{\JPA}
% Comment out if separate title page not required
\maketitle

\section{Necessary preliminaries}
There is growing interest in the study of non-Hermitian Hamiltonian operators due to the fact that, though non-Hermitian in nature, these operators may constitute valid quantum mechanical systems \cite{SGH92,HN96,BB98,Mo02}. In ref.\cite{SGH92}, reality of the spectrum of a set of non-Hermitian Hamiltonians within the context of a consistent quantum mechanical framework is linked to the existence of a positive definite and bounded metric operator, while in \cite{BB98} it is conjectured from numerical evidence that the Schr\"{o}dinger equation with the complex potential $V(x)=i x^3$ has real eigenvalues. Later a large number of papers \cite{LZ00} were published by various groups on non-Hermitian ${\cal{PT}}$-symmetric Hamiltonians and it is now established that  non-Hermitian Hamiltonians having ${\cal{PT}}$-symmetry have real eigenvalues if the corresponding eigenfunctions are also ${\cal{PT}}$ symmetric, otherwise the eigenvalues will occur in complex conjugate pairs (see \cite{Be07} for a recent survey on ${\cal{PT}}$-symmetry and additional references). Later in a series of papers \cite{Mo02}, it has been clarified that a non-Hermitian Hamiltonian having all eigenvalues real is connected to its Hermitian conjugate through a linear, Hermitian, invertible and bounded operator $\zeta : {\cal{H}} \rightarrow {\cal{H}}$ with bounded inverse, satisfying
\begin{equation}
H^\dag = \zeta H \zeta^{-1}
\end{equation}
i.e., $H$ is Hermitian with respect to a positive definite inner product defined by $\langle\langle.,.\rangle\rangle = \langle.,\zeta.\rangle$ and termed as $\zeta$-pseudo-Hermitian. Consequently one may define an invertible operator, $\rho = \sqrt{\zeta}$ which enables us to write \cite{SGH92,MB04,KS04}
 \begin{equation}
 h= \rho H \rho^{-1}\label{e2}
 \end{equation}
 where $h$ is equivalent Hermitian analogue of $H$ with respect to the standard inner product $\langle.,.\rangle$. However, $\rho$ need not be the be square root of a metric operator \cite{SGH92,KS04}.

The harmonic oscillator Hamiltonian augmented by a non-Hermitian ${\cal{PT}}$-symmetric part is one of the model falling into the class mentioned above. This non-Hermitian oscillator was first discussed by Swanson \cite{Sw04} who considered the Hamiltonian
 \begin{equation}
 H_{S} = w \left(a^{\dagger} a + \frac{1}{2}\right) + \alpha a^2 + \beta a^{\dagger^2},~~~~w, \alpha, \beta \in \mathbf{R}\label{e1}
 \end{equation}
 where $a, a^\dag$ are bosonic harmonic oscillator annihilation and creation operators satisfying usual commutation relationship $[a,a^\dag] = 1$. If $\alpha \ne \beta$, $H_S$ is non-Hermitian ${\cal{PT}}$-symmetric (${\cal{P}}$-pseudo-Hermitian). Swanson found a Bogoliubov type transformation
\begin{equation}
\left(
  \begin{array}{c}
    d \\
    c \\
  \end{array}
\right) = \left(
            \begin{array}{cc}
              g_4 & -g_2 \\
              -g_3 & g_1 \\
            \end{array}
          \right)
\left(
                    \begin{array}{c}
                      a \\
                      a^\dag \\
                    \end{array}
                  \right)
\end{equation}
with $c= g_1 a^\dag -g_3 a$ and $d= g_4 a- g_2 a^\dag.$ The algebra $[a,a^\dag] = 1$ gives only constraint on the operators $c$ and $d$ as $[d,c] = g_1 g_4 - g_2 g_3 = 1.$ It is not necessary that $c,d$ will be Hermitian conjugate. Swanson showed that the transformed Hamiltonian has the eigenvalue of a harmonic oscillator system with frequency $\sqrt{w^2-4\alpha \beta}$ so long as $w>\alpha+\beta$ and the eigen functions can be derived from the eigenfunctions of the harmonic oscillator.

Subsequently extensive studies related to the various aspects and generalization of the Hamiltonian (\ref{e1})
have been reported in \cite{Jo05,BQR05,MGH07,Qu07,BT08,SR07,SR09}. Jones \cite{Jo05},
showed that the non-Hermitian oscillator Hamiltonian $H_S$ given in (\ref{e1}) can be mapped to harmonic oscillator
Hamiltonian viz.,
\begin{equation}
h =\rho H_S~ \rho^{-1} = -\frac{1}{2} (w-\alpha-\beta)\frac{d^2}{dx^2} + \frac{1}{2} \frac{w^2-4 \alpha \beta}{w-\alpha -\beta} x^2
\end{equation}
 with $\rho = \exp[-\frac{1}{2}\frac{\alpha-\beta}{w-\alpha-\beta}x^2]$. Bagchi et al. \cite{BQR05}, explained the hidden symmetry of Swanson
Hamiltonian to find the operators $\zeta$ in the form $\zeta= \rho^{-1}_{(\alpha,\beta)} \rho_{(\beta,\alpha)}$.
The positivity of $\zeta$ is provided by the property satisfied by $\rho$ viz., $\rho_{(\alpha,\beta)} = \rho^{-1}_{(\beta,\alpha)}$. A word of caution is due here. This is because if one writes $\zeta= \rho^{-1}_{(\alpha,\beta)} \rho_{(\beta,\alpha)}$ then the boundedness of the operator $\zeta$ can not be guaranteed. Though in our case $\rho$ is different from the problem discussed in \cite{Jo05} because of the generalized form of $a$ and $a^\dag$, this problem still remains a serious matter of concern. If $\zeta$ is not bounded metric operator an alternative construction of $\rho$ can be made which is not based on the metric operator \cite{KS04}. However our main objective will be fulfilled by finding an invertible operator $\rho$ satisfying $h=\rho H \rho^{-1}$.

In ref.\cite{BQR05}, it has also been proposed a general first order differential form of the annihilation and creation operators $\tilde{a}, \tilde{a}^\dag$ viz.,
\begin{equation}
\tilde{a}= A(x) \frac{d}{dx} + B(x)~~~~\tilde{a}^\dag = -A(x) \frac{d}{dx} + B(x) -A'(x)\label{e3}
\end{equation}
respectively, where $A(x), B(x) \in \mathbf{R}$ and `prime' denotes derivative with respect to $x$. In this case $[\tilde{a},\tilde{a}^\dag] = 2A B'-A A''$ ($\ne 1$ in general). For this generalization the ${\cal{PT}}$-symmetry of the generalized Swanson Hamiltonian
\begin{equation}
H_{GS} = w \left(\tilde{a}^{\dagger} \tilde{a} + \frac{1}{2}\right) + \alpha \tilde{a}^2 + \beta \tilde{a}^{\dagger^2}
\end{equation}
is lost (unless $A(x)$ is an odd
function and $B(x)$ is an even function of x). For the general choice of $\tilde{a},\tilde{a}^\dag$ and $w-\alpha-\beta = 1$, the Hermitian equivalent form of the generalized Swanson Hamiltonian is obtained by a similarity transformation \cite{BQR05}
\begin{equation}
\tilde{h} = \rho_{(\alpha,\beta)} H_{GS}~ \rho_{(\alpha,\beta)}^{-1} = -\frac{d}{dx} A(x)^2 \frac{d}{dx}
+V_{eff} (x)\label{e6}
\end{equation}
where
\begin{equation}
\tilde{\rho}_{(\alpha,\beta)} = A(x)^{\frac{\alpha-\beta}{2}}
\exp\left(-(\alpha-\beta)\int^x \frac{B(x)}{A(x)}
dx'\right),\label{e7}
\end{equation}
\begin{equation}
\begin{array}{ll}
\displaystyle V_{eff}(x) = \frac{(\alpha+\beta) A A''}{2} + \left[\frac{\alpha+\beta}{2} + \frac{(\alpha-\beta)^2}{4}\right] A'^2 - 4 \tilde{w}^2 A'B \\
 \quad \quad \quad \quad \quad \displaystyle + 4 \tilde{w}^2 B^2 - (\alpha+\beta+1) A B' + \frac{\alpha+\beta+1}{2}\label{e15}
\end{array}
\end{equation}
\begin{equation}
\tilde{w} = \frac{\sqrt{1+ 2 (\alpha+\beta) + (\alpha-\beta)^2}}{2}.
\end{equation}
 There is an one to one correspondence between the energy eigenvalues of $\tilde{h}$ given in (\ref{e6}) and $H_{GS}$. If $\psi_n(x)$ are the wave functions of the equivalent Hermitian Hamiltonian $\tilde{h}$ then the wave functions of the Hamiltonian $H_{GS}$ are given by $\rho_{(\alpha,\beta)}^{-1} \psi_n(x)$.

The purpose of the present note is to study the generalized Swanson Hamiltonian $H_{GS}$ for the general form of creation and annihilation operators $\tilde{a}^\dag,\tilde{a}$ given in (\ref{e3}) for which $[\tilde{a},\tilde{a}^\dag] = 1$. In fact our analysis would hold if $[\tilde{a},\tilde{a}^\dag] = k$, $k\in\mathbf{R}$ is any constant. We will show that in this case the equivalent Hermitian Hamiltonian $\tilde{h}$ of the generalized Swanson Hamiltonian can be transformed into the harmonic oscillator like problem irrespective of the choice of $A(x)$. We will show that even in the case where the quantum condition $[\tilde{a},\tilde{a}^\dag] = 1$ is satisfied the generalized Swanson Hamiltonian is not necessarily isospectral to the harmonic oscillator. This apparent anomaly occurs because the domain of $x$ and the domain of the variable used in co-ordinate transformation to obtain the harmonic oscillator are not necessarily same. Our findings has been illustrated in the framework of position dependent mass scenario.

\section{Non-isospectrality of generalized Swanson Hamiltonian and harmonic oscillator}
\subsection{Reduction of equivalent Hermitian Hamiltonian $\tilde{h}$ into harmonic oscillator Hamiltonian}
Let us consider the case when commutator of $\tilde{a}$ and $\tilde{a}^{\dagger}$ is given by
\begin{equation}
[\tilde{a},\tilde{a}^\dagger] = 2 A(x) B'(x) - A(x) A''(x) = 1.
\end{equation}
On setting
\begin{equation}
z(x) = \int^x \frac{dx'}{A(x')}~~~~~~~~~~~~ B(x) =\frac{z''}{2z'^2} + \frac{1}{2} z, \label{e4}
\end{equation}
(for the sake of simplicity we assume integration constant to be zero), the Hamiltonian $\tilde{h}$ given in eq.(\ref{e6}) becomes
\begin{equation}
\tilde{h} = -\frac{d}{dx}A^2\frac{d}{dx} + V_{eff}(x),~~~~~V_{eff}(x) = \frac{1}{2} \frac{z'''}{z'^3} - \frac{5}{4} \frac{z''^2}{z'^4} + \tilde{w}^2 z ^2.\label{e5}
\end{equation}
The important point is to note here that the co-ordinate transformation given by the first of the equation (\ref{e4}) has certain peculiarities. The domain of $z(x)$ depends on the functional form of $A(x)$ and may not be the same as that of $x$. This may lead as will be shown by concrete examples in section 3, to non-isospectrality of the original Hamiltonian $\tilde{h}$ and the transformed Hamiltonian. The examples are discussed in the framework of position dependent mass Schr\"{o}dinger system where $m(x)$, the position dependent mass is identified with $1/A(x)^2$.

For the change of variable (\ref{e4}) we have
\begin{equation*}
A'= \frac{\dot{A}}{A},~~~~~~~ A'' = \frac{\ddot{A}}{A^2} -
\frac{\dot{A}^2}{A
^3}~~~~~ \mbox{etc.}
\end{equation*}
where `dot' represents derivative with respect to $z$.
Consequently the Schr\"{o}dinger eigenvalue equation for the
Hamiltonian $\tilde{h}$ reduces to
\begin{equation}
\left[-\frac{d^2}{dz^2} -\frac{\dot{A}}{A} \frac{d}{dz} +
\left(\frac{\dot{A}^2}{4A^2}-\frac{\ddot{A}}{2 A} + \tilde{w}^2 z
^2\right)\right]\psi(z) = E \psi(z)\label{e8}
\end{equation}
In order to eliminate the first derivative term from equation
(\ref{e8}) we assume
\begin{equation}
\psi(z) = A(z)^{-\frac{1}{2}} \phi(z)\label{e9}.
\end{equation}
For this assumption equation (\ref{e8}) reduces to
\begin{equation}
\left[-\frac{d^2}{dz^2} + \tilde{w}^2 z^2\right] \phi(z) = E \phi(z)\label{e10}
\end{equation}
This equation (\ref{e10}) is the Schr\"{o}dinger equation with harmonic oscillator potential $V(z) = \tilde{w}^2 z^2$ and can be transformed into confluent hypergeometric equation
\begin{equation}
y \frac{d^2\chi}{dy^2} +\left(\frac{1}{2}-y\right) \frac{d\chi}{dy} + \left(\frac{E}{4\tilde{w}}-\frac{1}{4}\right) \chi =0\label{e12}
\end{equation}
by the following transformations
\begin{equation}
y= \tilde{w} z^2~~~~~~~~~~~~~ \phi(y) = e^{-\frac{y}{2}} \chi(y).\label{e11}
\end{equation}
Hence the general solution of the equation (\ref{e10}) are given by
\begin{equation}
\phi_{e} (z) \sim
e^{-\frac{\tilde{w}}{2} z^2} {_1}F_1
\left(\frac{1}{4}-\frac{E}{4 \tilde{w}},\frac{1}{2},\tilde{w}z^2\right)\label{e13}
\end{equation}
\begin{equation}
\phi_{o} (z) \sim
z e^{-\frac{\tilde{w}}{2} z^2} {_1}F_1
\left(\frac{3}{4}-\frac{E}{4 \tilde{w}},\frac{3}{2},\tilde{w}z^2\right),\label{e14}
\end{equation}
where $\phi_e$ and $\phi_o$ denote the even and odd solutions. The eigen functions of the Hamiltonian $\tilde{h}$ are given by
\begin{equation}
\psi(x)\sim A(z)^{-{\frac{1}{2}}} \phi(z),
\end{equation}
where $z(x)$ is given by equation (\ref{e4}).
\subsection{Condition of Isospectrality}
Now one obtains the eigenvalues of the equation (\ref{e10}) by analyzing the behaviors of the eigenfunctions (\ref{e13}) and (\ref{e14}). If the domain of argument $\tilde{w} z^2$ is unbounded then $\phi_e(z)$  and $\phi_o(z)$ will not in general, square integrable because of the asymptotic behavior of the confluent Hypergeometric function viz. \cite{AS70},
\begin{equation}
_1F_1(a,b,y) = \frac{\Gamma(b)}{\Gamma(a)} e^y y^{a-b} [1+O(y^{-1})],~~~~Re(y)>0.
\end{equation}
So in order to make the eigenfunctions square integrable one must take $a=-m$ $(m=0,1,2,3...)$ in which case $_1F_1(a,b,y)$ reduces to a polynomial. So from (\ref{e13}) $1/4 - E/(4 \tilde{w}) = -m$ or equivalently
\begin{equation}
E_{2m} = 2\tilde{w}\left(2 m + \frac{1}{2}\right)\label{e16}
\end{equation}
and from (\ref{e14}) $3/4-E/(4 \tilde{w}) = -m$ or equivalently
\begin{equation}
E_{2m+1}= 2\tilde{w}\left(2m+\frac{3}{2}\right)\label{e17}
\end{equation}
Combining (\ref{e16}) and (\ref{e17}) we have
\begin{equation}
E_n = 2 \tilde{w} \left(n + \frac{1}{2}\right),~~ n=0,1,2...
\end{equation}
Hence in this case the Hamiltonian $\tilde{h}$ has the spectrum of a harmonic oscillator.

However if the domain of $\tilde{w}z^2$ is finite then the required boundary condition to be satisfied by the eigenfunctions (\ref{e13}) and (\ref{e14}) is that they must vanish at the end points of the domain of $z$ and the eigenvalues are given by the zeroes of the confluent hypergeometric functions when the arguments attain their end points. The first approximation of the $m$'th ($m=1,2...$) positive zero $X_0$ of ${_1}F_1(a,b,y)$ is given by \cite{AS70}
\begin{equation}
X_0 = \frac{\pi^2\left(m+\frac{b}{2}-\frac{3}{4}\right)^2}{2 b -4 a} \left[1+ O\left(\frac{1}{(\frac{b}{2}-a)^2}\right)\right],~~~m=1,2...
\end{equation}
Hence from eqns.(\ref{e13}) and (\ref{e14}) we obtain
\begin{equation}
E_{2m}\approx \frac{\pi^2}{4 z_{\pm}^2} (2m-1)^2,~~~E_{2m-1} \approx \frac{\pi^2}{4 z_{\pm}^2} (2m)^2
\end{equation}
respectively. Combining these two we obtain
\begin{equation}
E_n \approx \frac{\pi^2}{4 z_{\pm}^2} n^2,~~n=1,2,3...\label{e18}
\end{equation}
where $z_{\pm}$ are the end points of the domain of definition of $V(z)$. Hence in this case the Hamiltonian $\tilde{h}$ given in (\ref{e5}) will not be isospectral to the harmonic oscillator.
\section{Connection with position-dependent mass models}
As mentioned before, the generalization of the Swanson model enables one to connect those physical systems which are describable by a position dependent mass by choosing $ A(x) = m(x)^{-1/2}$ which is a strictly positive function. In this case the Hamiltonian (\ref{e6}) reduces to the time independent position-dependent mass Hamiltonian \cite{BD66,Le95} for which the Schr\"{o}dinger equation reeds
\begin{equation}
\left(-\frac{d}{dx} \frac{1}{m(x)}\frac{d}{dx} + V_{eff} (x)
\right)\psi(x) = E \psi(x)
\end{equation}
with $V_{eff}(x)$ is given by equation (\ref{e5}).
In the following we shall considere two different mass profile $m(x)$ to illustrate that the Hamiltonian given in (\ref{e6}) is not always isospectral to harmonic oscillator.

\subsection{Isospectral case}
Let us consider the following mass function
\begin{equation}
m(x) = \frac{1}{1+x^2}~,~~x\in (-\infty,\infty)
\end{equation}
which has been considered in the study of quantum nonlinear
oscillator \cite{MR09,Ca07}. For this choice of mass function, $z(x)$ is
given by
\begin{equation}
z(x) = sinh^{-1} (x).
\end{equation}
It is clear that $z(x)\rightarrow \pm \infty$ as $x\rightarrow \pm\infty$.
So in this case, we have the Hamiltonian $\tilde{h}$ with the effective potential $V_{eff}(x)$
\begin{equation}
V_{eff} (x) = -\frac{2+x^2}{4(1+x^2)} + \tilde{w}^2 (sinh^{-1}~ x)^2
\end{equation}
is isospectral to harmonic oscillator.
In table 1, we have given a list of physically interesting mass functions, $V_{eff}(x)$, and eigen energies, for which the corresponding position dependent mass Hamiltonians are isospectral to the harmonic oscillator.
\subsection{Non-isospectral case}
Now let us choose
\begin{equation}
m(x) = sech^2(x),~~~x\in (-\infty,\infty)
\end{equation}
which depicts the solitonic profile \cite{Ba07}. For this choice, the function
$z(x)$ reeds
\begin{equation}
z(x) = tan^{-1}(sinh~ x).
\end{equation}
Here as $x\rightarrow\pm \infty, z(x)\rightarrow \pm \pi/2$, so the eigenvalues  of the Hamiltonian $\tilde{h}$ with
\begin{equation}
V_{eff}(x) = \frac{1}{4} - \frac{3}{4} cosh^2~x +
\tilde{w}^2\left(tan^{-1} (sinh~x)\right)^2,
\end{equation}
will be given by zeroes of the functions given in eqns.(\ref{e13}) and (\ref{e14}) at $z=\pm \pi/2$. First approximate value of the energy eigenvalues are given by equation (\ref{e18})
\begin{equation}
E_n \approx n^2, ~n=1,2,3...
\end{equation}
In table 2, we have given a list of physically interesting mass functions, $V_{eff}(x)$, and eigen energies, for which the corresponding Hamiltonians $\tilde{h}$ are not isospectral to the harmonic oscillator.

\begin{table}
\begin{center}
\begin{tabular}{|c|c|c|}
  \hline
  % after \\: \hline or \cline{col1-col2} \cline{col3-col4} ...
  $m(x)$ & $z(x)$ &$V_{eff}(x)$ \\
  \hline
  \hline
   $\frac{1}{(1+x^2)}$&$sinh^{-1}~x$  & $-\frac{2+x^2}{4(1+x^2)} + \tilde{w}^2 (sinh^{-1}~ x)^2$ \\
   \hline
   $\cosh^2 x$&$sinh~x$  & $\frac{1}{8} (7-3\cosh 2x) sech^4~x + \tilde{w}^2 sinh^2~x$  \\
   \hline
   $\left(\frac{\gamma+x^2}{1+x^2}\right)^2$&$x+(\gamma-1)\tan^{-1}~x$  & $ \frac{(\gamma-1)(3x^4-2(a-2)x^2-a)}{(x^2+\gamma)^4}+\tilde{w}^2 [x+(\gamma-1)\tan^{-1}~x]^2$  \\
   \hline
   $e^{2 x} sech^2~x$&$\log (1+e^{2 x})$  & $-\frac{3}{4} e^{-4 x}-\frac{1}{2}e^{-2 x}+\tilde{w}^2 [\log (1+e^{2 x})]^2  $   \\
   \hline
   $e^{-x}$& $-2e^{-\frac{x}{2}}$&$-\frac{3}{16} e^x+4 \tilde{w}^2 e^{-x}$\\
  \hline
  \end{tabular}
  \caption{Isospectral case: some physically interesting mass functions and corresponding effective potentials for which the corresponding Hamiltonians $\tilde{h}$ are isospectral to harmonic oscillator with $E_n=2\tilde{w}\left(n+\frac{1}{2}\right),n=0,1,2...$. In all these cases $z(x)$ are unbounded as $x\rightarrow\pm \infty.$ }
  \end{center}
  \end{table}
 
  \begin{table}
  \begin{center}
  \begin{tabular}{|c|c|c|c|}
  \hline
  $m(x)$ & $z(x)$ &$V_{eff}(x)$ & $E_n \approx$\\
  \hline
  \hline
  $sech^2 x$& $\tan^{-1}(sinh~x)$&$\frac{1}{4}-\frac{3}{4}\cosh^2 x +\tilde{w}^2 (\tan^{-1}(sinh~x))^2$&$n^2$\\
  \hline
  $e^{-2x^2}$&$\frac{\sqrt{\pi}}{2} Erf~ x$&$-(1+3x^2)e^{2 x^2}+\frac{\pi\tilde{w}^2}{4} (Erf~ x)^2$ &$\pi n^2$\\
  \hline
  $\frac{1}{(1+x^2)^2}$& $\tan^{-1} ~x$ & $-(1+2 x^2)+\tilde{w}^2 (\tan^{-1}~x)^2$&$n^2$\\
  \hline
  \end{tabular}
  
  \caption{Non-isospectral case: some physically interesting mass functions and corresponding effective potentials for which the corresponding Hamiltonians $\tilde{h}$ are not isospectral to harmonic oscillator. In all these cases $z(x)$ are finite as $x\rightarrow\pm \infty.$ Here $n$ takes the values $1,2,3...$ }
  \end{center}
  \end{table}
  
\section{Conclusion}
To conclude, we have studied a class of non-Hermitian Hamiltonians of the form $H_{GS} = w(\tilde{a}\tilde{a}^\dag + 1/2) +\alpha \tilde{a}^2 + \beta \tilde{a}^{\dag^2}$, where $w,\alpha,\beta$ are real constants and $\tilde{a},\tilde{a}^\dag$ are generalized annihilation and creation operators. For the constraint $[\tilde{a},\tilde{a}^\dag]=constant$, the equivalent Hermitian Hamiltonian $\tilde{h}$ of $H_{GS}$ can be transformed into what looks like a harmonic oscillator problem, but it has been shown that $\tilde{h}$ is not always isospectral to harmonic oscillator because of the peculiarities of the transformation. Our findings has been explained in the framework of position dependent mass model. We hope that the non-isospectrality problem discussed here could occur in non-Hermitian Hamiltonians other than the generalized Swanson model and hence this paper may lead to more studies on this direction.

\end{document}